\documentclass[twoside,a4paper,11pt]{sea22}
\usepackage{graphicx}
\usepackage{hyperref}
\usepackage{media9}
\begin{document}
\pagenumbering{arabic}
\pagestyle{myheadings}
\thispagestyle{empty}
{\flushleft\includegraphics[width=\textwidth,bb=58 650 590 680]{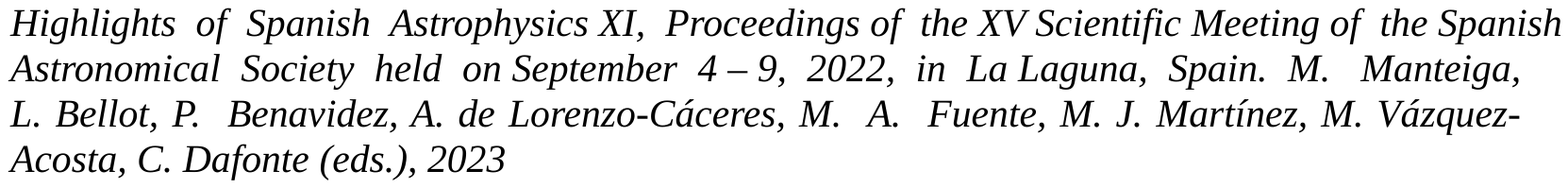}}
\vspace*{0.2cm}
\begin{flushleft}
{\bf {\LARGE
%
%%% TITLE of the paper. 
%%% TITLE of the paper. 
Web-based telluric correction made in Spain: spectral fitting of Vega-type telluric standards
%
% Do not delete next few lines
}\\
\vspace*{1cm}
%
%%% Include here the LIST OF AUTHORS.
%%% Include here the LIST OF AUTHORS.
%%% Note that the last author has to be preceeded by an AND.
de la Fuente, D.$^{1,2}$,
Marco, A.$^{1}$,
Patrick, L.~R.$^{1,3,4}$, 
R\"ubke, K.$^{1}$,
L\'opez, I.$^{1}$,
Fern\'andez, A.$^{1}$,
Conejero, S.$^{1}$,
Navarro, J.$^{1}$,
Palaz\'on, M.$^{1}$,
and
Negueruela, I.$^{4}$
%
% Do not delete next few lines
}\\
\vspace*{0.5cm}
%
%%% AFFILIATIONS LIST.
%%% and the AFFILIATIONS LIST. Note that one affiliation per line.
%%% Add as many affiliations as necessary. 
$^{1}$Departamento de F\'isica, Ingenier\'ia de Sistemas y Teor\'ia de la Se\~nal, Universidad de Alicante, San Vicente del Raspeig, Spain\\
$^{2}$CREOL, The College of Optics and Photonics, University of Central Florida, Orlando, FL, USA\\
$^{3}$Departamento de Astrof\'isica, Centro de Astrobiolog\'ia (CSIC-INTA), Torrej\'on de Ardoz, Spain\\
$^{4}$Departamento de F\'isica Aplicada, Universidad de Alicante, San Vicente del Raspeig, Spain\\
%
% Do not delete next few lines
\end{flushleft}
%
% Headings
\markboth{
%%% Type the SHORT version of the paper title.
%%% Type the SHORT version of the paper title.
Web-based telluric correction through standard stars
}{ % Do not delete
%
%%%  First Author \& Second Author   OR   First-author et al. 
%%%  First Author \& Second Author   OR   First-author et al. if the author list 
%%% contains three or more authors.
de la Fuente et al.
% 
% Do not delete next few lines
}
\thispagestyle{empty}
\vspace*{0.4cm}
\begin{minipage}[l]{0.09\textwidth}
\ 
\end{minipage}
\begin{minipage}[r]{0.9\textwidth}
\vspace{1cm}
\section*{Abstract}{\small
%
% ABSTRACT ABSTRACT ABSTRACT
% ABSTRACT ABSTRACT ABSTRACT
%%% Type the ABSTRACT of your oral contribution or poster
Infrared spectroscopic observations from the ground must be corrected from telluric contamination to make them ready for scientific analyses. However, telluric correction is often a tedious process that requires significant expertise to yield accurate results in a reasonable time frame. To solve these inconveniences, we present a new method for telluric correction that employs a roughly simultaneous observation of a Vega analog to measure atmospheric transmission. After continuum reconstruction and spectral fitting, the stellar features are removed from the observed Vega-type spectrum and the result is used for cancelling telluric absorption features on science spectra. This method is implemented as \texttt{TelCorAl} (Telluric Correction from Alicante), a Python-based web application with a user-friendly interface, whose beta version will be released soon.
%
% Do not delete next few lines
\normalsize}
\end{minipage}
%
%
%%% BODY of the paper
%%% BODY of the paper
%
\section{Introduction \label{intro}}

Infrared spectra that are observed from ground-based facilities are heavily contaminated by absorption features from the Earth's atmosphere. The accuracy of the decontamination process, known as telluric correction, is crucial for producing reliable results from subsequent spectroscopic analyses.

Telluric correction basically consists in dividing the science spectra by atmospheric transmission. However, the latter cannot be easily determined because the atmosphere behaves in a complex and unpredictable manner. The intensities and shapes of telluric absorptions depend not only on observational settings (airmass, instrumental profile, the observer's altitude) but also on atmospheric properties that can change dramatically (abundances for the involved molecules, precipitable water vapor, seeing).

The currently available methods mainly differ on how this transmission spectrum is obtained. On the one hand, theoretical methods (e.g. \cite{smette+15}) fit a transmission model to the observations. On the other hand, observational methods (e.g. \cite{vacca+03}) extract the transmission spectrum from a roughly simultaneous observation of a standard star whose intrinsic features are well known and easily separated from telluric absorptions. There also are some hybrid approaches (e.g. \cite{cook+22}). The user often has to make a complex choice, based on instrumental and operational features (spectral resolution, exposure times, observation schedule, etc.), and other requirements related to the planned treatment (intended signal-to-noise ratio, integration into a pipeline, etc.).

The majority of publicly available algorithms are designed for specific instruments or tailored to the scientific goals of some particular team. A few others have a more general purpose and are targeted to a wider community, although they have practical disadvantages in common. For example, users often complain about problems for installation and usability of these software, or about the amount of experience needed to obtain accurate results in a reasonable amount of time.

This situation has encouraged us to create a cross-platform, user-friendly application for telluric correction. To avoid the complications of fitting sky models or empirical libraries to the telluric component of science spectra, we have chosen the approach of measuring atmospheric transmission on an Observed Standard Star Spectrum (OSSS). We refer to \cite{sameshima+18b} and \cite{vacca+03} for the theoretical background for this kind of methods.

\section{Methodology} \label{sec:method}

Vega-type standards (i.e. spectral types around A0V) are preferred, as their intrinsic spectra only display wide hydrogen absorptions in most cases (see \cite{sameshima+18a} for a notable counterexample). These lines are usually well described by Voigt functions (see \cite{posener59} and references therein), which facilitates the modeling and removal of the OSSS stellar features. Model fitting, however, cannot be directly performed within wavelength regions where the continuum disappears from the observation (owing to broad telluric bands or to overlapping high-order hydrogen lines).

The extraction of the atmospheric transmission spectrum in such a way is the first phase of telluric correction, and is carried out in five steps:

\begin{figure}[!t]
\center
\includegraphics[width=15cm]{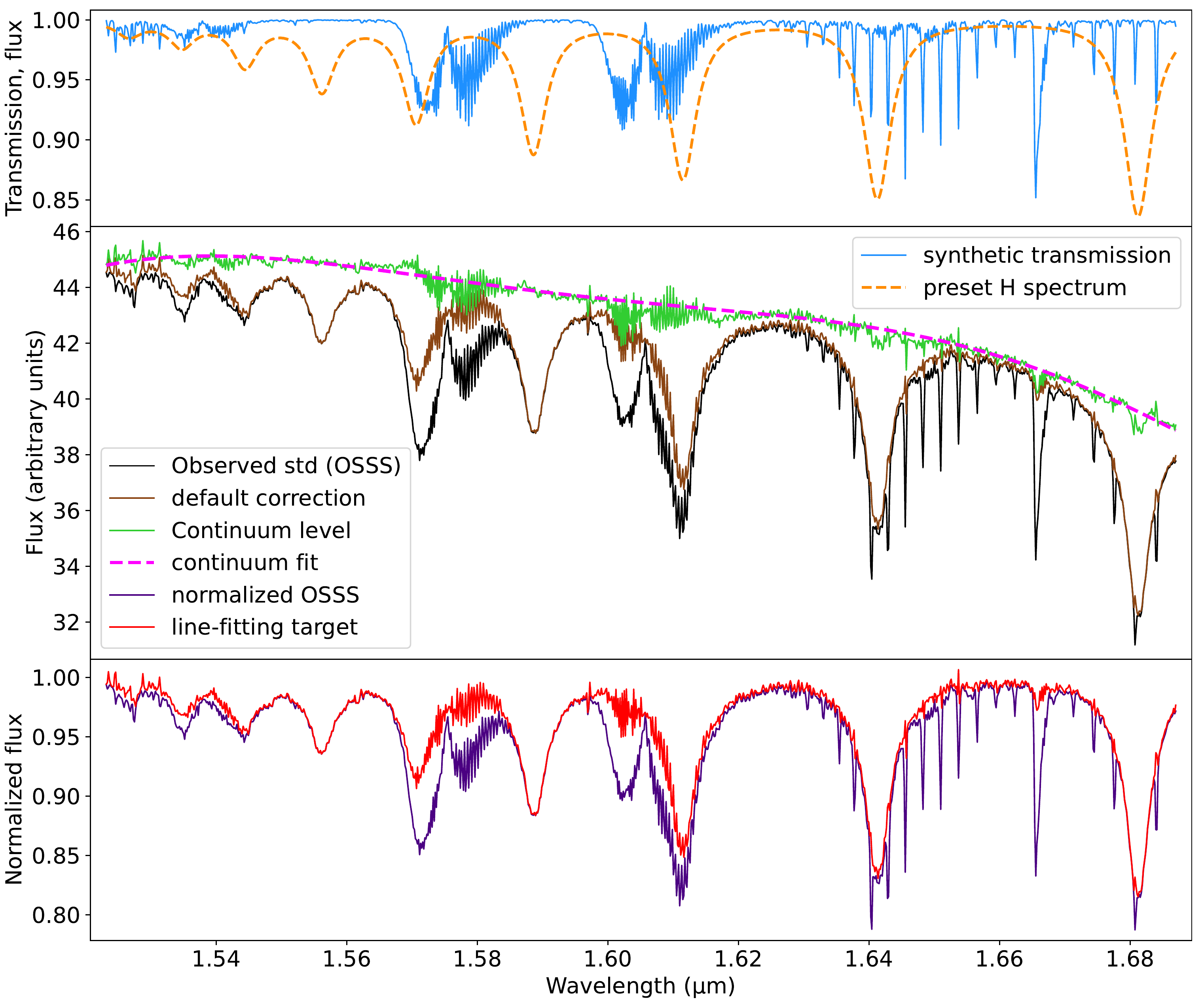} 
\caption{\label{fig:std} Spectra involved in the process of fitting a spectrum of a telluric standard star, as well as the normalized version of the OSSS for comparison purposes.}
\end{figure}

\begin{figure}[!t]
\center
\includegraphics[width=11.7cm]{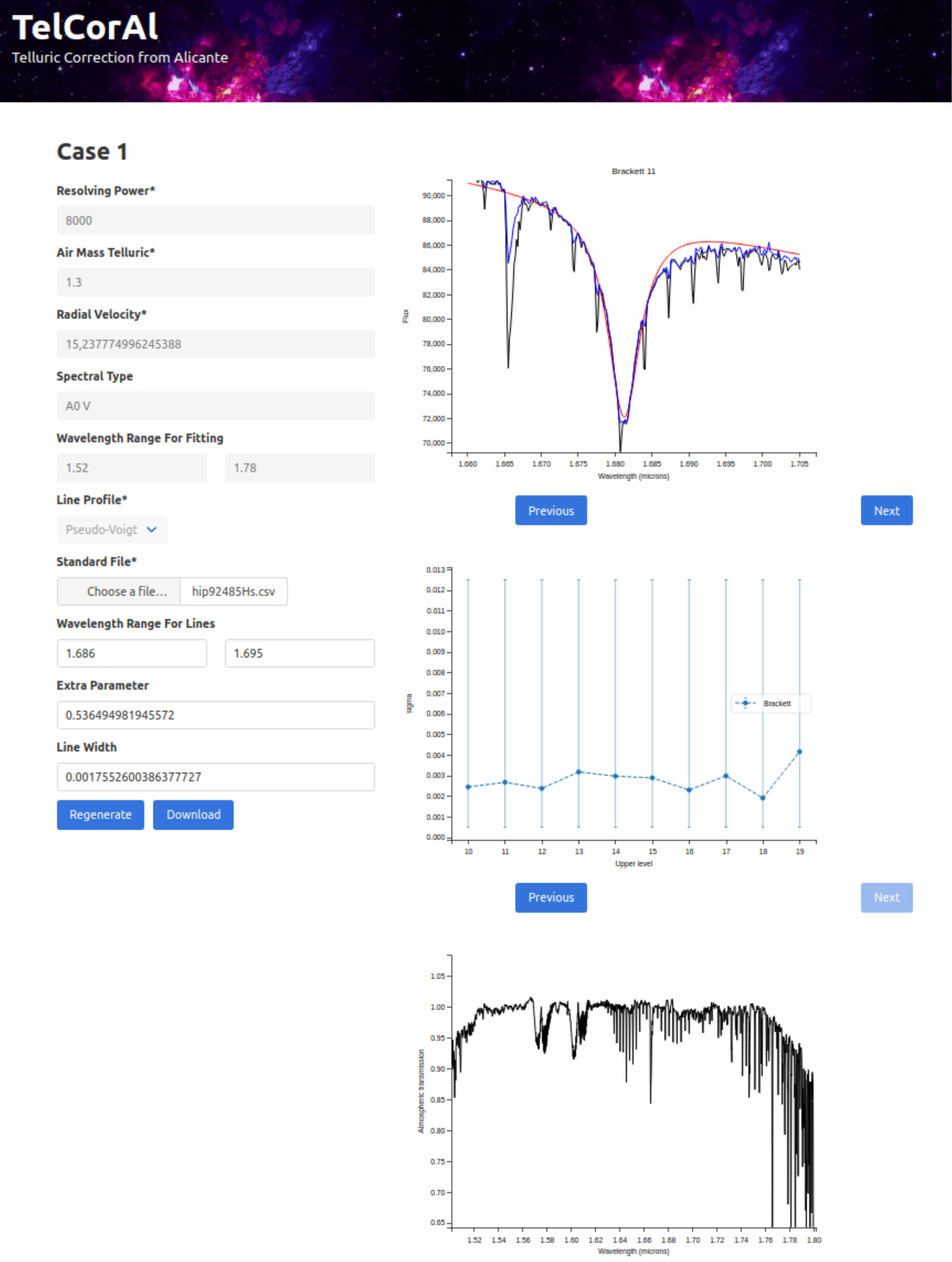} 
\caption{\label{fig:telcoral} \texttt{TelCorAl} interface, showing an example of the fitting process for intrinsic features of a telluric standard star.}
\end{figure}

\begin{enumerate}

  \item A ``default telluric correction'' of the standard star is made by dividing the OSSS by a synthetic trasmission spectrum (respectively shown in black and light blue in Fig. \ref{fig:std}). To generate the latter, the \cite{moehler+14} telluric models for average atmospheric conditions are interpolated at the airmass of the OSSS, convolved with a Gaussian profile that matches the spectral resolution of the instrument, and resampled into the OSSS wavelength axis. The \cite{moehler+14} models also provide variability information that is used for computing the uncertainty of the default correction.
  
  \item A simple model of hydrogen absorption spectrum is created as a combination of Voigt profiles whose widths and relative intensities are preset for the spectral type of the OSSS. The profile heights are then multiplied by a scaling constant to match the line heights of the default telluric correction (brown line in Fig. \ref{fig:std}), thus obtaining a first model of the standard star (dashed orange line).
  
  \item The ratio between the results of previous steps (i.e. brown line divided by orange line) is used as an approximation for the continuum level (shown in green in Fig. \ref{fig:std}). Weighted least-squares polynomial fits with boundary conditions are performed for increasing degrees of the polynomial, and the optimal result (magenta dashed line) is chosen on the basis of a goodness-of-fit test.
  
  \item The spectrum that results from the default correction is normalized (i.e. brown/magenta = red in Fig. \ref{fig:std}) and used for fitting a model for the combination of all H line profiles. The result is multiplied by the continuum fit (magenta dashed line) to produce a full model for the intrinsic features of the OSSS (See upper graph in Fig. \ref{fig:telcoral}). 

  \item The telluric transmission spectrum (Fig. \ref{fig:telcoral}, lower plot) is obtained as the ratio of the OSSS to its full model.
  
\end{enumerate}

If one or both of the initial synthetic spectra (like those in Fig. \ref{fig:std}, top pannel) diverges significantly from the OSSS features, the fitting procedures will produce inaccurate results. In this case, the method will iterate on steps 1--5 so that the model and transmission results of an iteration are used as initial spectra for the next one.

Once an accurate result for telluric transmission is achieved, the second phase consists in using the resulting spectrum for correcting the science spectra. In an ideal case, the telluric features imprinted on all these spectra would be identical, and could be perfectly removed by dividing each science spectrum by the transmission. In practice, however, there are intensity and shape variations, caused not only by variability of the sky (for non-simultaneous observations), but also by factors related with the reduction process (e.g. wavelength calibrations of co-added frames) or with implicit approximations (e.g. the definition of airmass). To deal with these differences, the following tasks are carried out for each science spectrum:

\begin{itemize}
 \item The transmission spectrum is slightly red- or blue-shifted according to a factor that corresponds to the maximum cross-correlation with the science spectrum. As far as possible, the cross-correlation function should be computed in a wavelength range where telluric features dominate over the intrinsic features of the science spectrum.
 
 \item The ratio between airmass factors for the science and transmission spectra is used for scaling the strength of telluric features through an exponential relation (\cite{perliski-solomon93}). The result can be manually fine-tuned by the user until optimal telluric cancellation is achieved.
\end{itemize}

\section{Implementation and current status}

We have developed \texttt{TelCorAl} (Telluric Correction from Alicante), a user-friendly web application that performs telluric correction in a semiautomatic way. Apart from carrying out all the telluric correction process as explained in Sect. \ref{sec:method}, the users may choose to perform just one phase, or both phases independently, in order to test the method or make the correction in a more customized way. In all cases, the users are required to upload their input spectra in comma-separated value format, and the results are delivered in the same format together with information on the parameters used throughout the process. All calculations are internally made by a \texttt{Python 3} code, and users interact with it through a web interface that also displays helpful plots (Fig. \ref{fig:telcoral}). Everything is processed in the RAM of the server, and no data or results are ever stored outside it, thus keeping them private at all times. 

The combined H line models are built and fitted through \texttt{lmfit 1.0.3} (\cite{newville+14,newville+21}), by employing a line profile function of the user's choice. Based on our experience, the pseudo-Voigt approximation (\cite{whiting68}) is usually the option that works better in general, although Gaussian or Lorentzian profiles are also offered, as they are more suitable under certain circunstances (e.g. extreme values for resolution and $v \sin i$ of the standard star). Other critical components of fitting and processing the spectra for telluric correction make use of the \texttt{SciPy} (\cite{virtanen+20}), \texttt{PyAstronomy} (\cite{czesla+19}), and \texttt{specutils} (\cite{earl+22}) libraries.

A preliminary version of \texttt{TelCorAl} is currently running on the server of our group, although being only accessible from the University of Alicante network for the moment. Once the final telluric correction code is fine-tuned and comprehensive testing is carried out, a beta version will be released to the public.

\section{Concluding remarks} 

Our forthcoming web application solves issues that make other telluric correction software impractical for widespread use. The user-friendly webpage format, which does not require installation, along with the semiautomatic approach, makes our software suitable even for inexperienced users that aim at promptly getting a decent correction of their infrared data. \texttt{TelCorAl} is also capable of dealing with a wide variety of instrumental and observational features of the input data.

Compared to other telluric correction methods that are based on spectral fitting of Vega-type stars (\cite{vacca+03,sameshima+18b}), our code uses a novel approach that consists in reconstructing the continuum level (green line in Fig. \ref{fig:std}). This allows us to achieve accurate fits on regions where hydrogen line profiles overlap with each other or with wide telluric bands (e.g. between 1.54 and $1.62~\mu\mathrm{m}$, see Fig. \ref{fig:std}).

%
%
% Do not delete the next line
\small  % Do not delete
%
%%% Comment the following line if you do not have acknowledgments.
\section*{Acknowledgments}   % Do not delete if you declare acknowledgments
%
%%% ACKNOWLEDGMENTS
%%% ACKNOWLEDGMENTS
This work has been financially supported by Generalitat Valenciana through grants APOSTD/2020/228, APOSTD/2020/247, and PROMETEO/2019/041. The Spanish Ministerio de Ciencia e Innovaci\'on and Agencia Estatal de Investigación (MCIN/AEI/10.13039/501100011033/FEDER, UE) has also provided financial support through grants PGC2018-093741-B-C21 and PID2021-122397NB-C22.
%
% Do not delete the next few lines

%

\begin{thebibliography}{}
\small
%
%%% BIBLIOGRAPHY
%%% BIBLIOGRAPHY

\bibitem{czesla+19} Czesla, S., Schr{\"o}ter, S., Schneider, C.~P., et al. 2019, ascl:1906.010

\bibitem{cook+22} Cook, N.~J., Artigau, {\'E}., Doyon, R., et al. 2022, arXiv:2211.01358

\bibitem{earl+22} Earl, N., Tollerud, E., Jones, C., et al. 2022, Zenodo

\bibitem{moehler+14} Moehler, S., Modigliani, A., Freudling, W., et al. 2014, A\&A 568, A9

\bibitem{newville+21}Newville, M., Otten, R., Nelson, A., et al. 2021, Zenodo

\bibitem{newville+14}Newville, M., Stensitzki, T., Allen, D.~B., et al. 2014, Zenodo

\bibitem{perliski-solomon93} Perliski, L.~M. \& Solomon, S. 1993, JGR, 98, 10363

\bibitem{posener59} Posener, D.~W. 1959, AuJPh., 12, 184

\bibitem{sameshima+18a} Sameshima, H., Ikeda, Y., Matsunaga, N., et al. 2018, ApJS, 239, 19

\bibitem{sameshima+18b} Sameshima, H., Matsunaga, N., Kobayashi, N., et al. 2018, PASP, 130, 074502

\bibitem{smette+15} Smette, A., Sana, H. Noll, S. et al. 2015, A\&A 576, A77

\bibitem{vacca+03} Vacca, W.~D., Cushing, M.~C., \& Rayner, J.~T. 2003, PASP, 115, 389

\bibitem{virtanen+20} Virtanen P., Gommers R., Oliphant T.~E., et al. 2020, NatMe, 17, 261

\bibitem{whiting68} Whiting, E.~E. 1968, JQSRT, 8, 1379

%
%
% Do not delete next few lines
\end{thebibliography}
\end{document}